\documentclass[12pt,preprint]{aastex}

\newcommand{\p}{\ensuremath{\mem{p}}}

\newcommand{\czw}{\ensuremath{^{12}\mem{C}}}

\newcommand{\nvi}{\ensuremath{^{14}\mem{N}}}

\newcommand{\ofu}{\ensuremath{^{15}\mem{O}}}
\newcommand{\ose}{\ensuremath{^{16}\mem{O}}}

\newcommand{\msun}{\ensuremath{\, {\rm M}_\odot}}

\newcommand{\beq}{\begin{equation}}
\newcommand{\beqa}{\begin{eqnarray}}
\newcommand{\eeq}{\end{equation}}
\newcommand{\eeqa}{\end{eqnarray}}
\newcommand{\bedis}{\begin{displaymath}}
\newcommand{\edis}{\end{displaymath}}

\newcommand{\mem}[1]{\ensuremath{\mathrm{ #1}}}

\newcommand{\punkt}{\mathrm{ \hspace*{0.1cm}.}}

\newcommand{\kap}[1]{Sect.\,\ref{#1}}

\newcommand{\abb}[1]{Fig.\,\ref{#1}}

\newcommand{\tab}[1]{Table\,\ref{#1}}

\usepackage{natbib}

\shorttitle{C-stars and Reaction Rates}
\shortauthors{Herwig & Austin}

\begin{document}

\title{Nuclear Reaction Rates and Carbon Star Formation}

\author{Falk Herwig}
\affil{Los Alamos National Laboratory, Theoretical Astrophysics
   Group T-6, MS B227, Los Alamos, NM 87545}
\email{fherwig@lanl.gov}
\author{Sam M. Austin}
\affil{National Superconducting Cyclotron Laboratory and Joint
Institute for Nuclear Astrophysics (JINA), Michigan State
University, East Lansing MI 48824}
\email{austin@nscl.msu.edu}

\begin{abstract}
  We have studied how the third dredge-up and the carbon star
  formation in low-mass Asymptotic Giant Branch stars depends on 
  certain  key nuclear reaction rates.  We find from a
  set of complete stellar evolution calculations of a $2\msun$ model
  with $Z=0.01$ including mass loss, that varying either the
  $\nvi(\p,\gamma)\ofu$ or the triple-$\alpha$ reaction rate
  within their uncertainties as given in the NACRE compilation results in dredge-up
  and yields that differ by a factor of 2. Model tracks with a
  higher rate for the triple-$\alpha$ rate and a lower rate for the
  $\nvi(\p,\gamma)\ofu$ reaction both show more efficient third
  dredge-up. New experimental results for the $\nvi(\p,\gamma)\ofu$
  reaction rates are surveyed, yielding a rate which is about $40\%$
  lower than the tabulated NACRE rate, and smaller than NACRE's lower
  limit. We discuss the possible implications of the revised nuclear reaction
  stellar evolution calculations that aim to reproduce the observed
  carbon star formation at low mass, which requires efficient third
  dredge-up.
\end{abstract}


\section{Introduction}
Low mass stars evolve through the core H- and He-burning stages to
finally enter into the Asymptotic Giant Branch (AGB) phase of
evolution \citep{iben:83b}. In this terminal phase most of the
nuclear processing that is relevant for  Galactic chemical
evolution is taking place. Two shells on top of the
electron-degenerate C/O core are burning H and He respectively.
Due to the different energy generation per mass unit the shells
have different Lagrangian speeds, and eventually shell burning
becomes unstable. He-shell flashes occur that prompt a series of
convective mixing episodes. First the region between the H- and
the He-shell becomes convectively unstable because of the large
energy generation in the thermonuclear runaway of the He-shell.
Then the layer at the bottom and below the convective envelope
expands and cools, which makes this region unstable against
convection. Layers previously covered by the He-shell flash
convection zone will become part of the convective envelope and
processed material from the core (or intershell to be more
precise) will be mixed to the surface. This process is called the
third dredge-up. Repeated dredge-up events will turn the initially
O-rich giant into a C-star with  $ \mem{C/O}>1$.

More than 20 years ago \citet{iben:81} pointed out that observed and theoretically predicted parameters of C-stars disagree. On the one
hand, models did predict C-star formation (i.e.\ efficient third
dredge-up) for large (core-)masses, which implies large stellar
luminosities. But no C-stars were observed at such high
luminosities. Instead,
observations showed that C-stars have low luminosities. But
stellar models of low mass were not able to reproduce the required
efficiency of dredge-up, or any dredge-up at all. In subsequent
years some progress has been made. The first disagreement 
has been resolved by the discovery that the envelope of massive
AGB stars is nuclear processed by CNO cycling because the bottom
of the envelope convection zone reaches into the H-shell
\citep{boothroyd:93}.

For the second disagreement concerning the low-mass stars no final solution has been agreed upon. The most
recent work by \citet{karakas:02a} gives a detailed account of the
dependence of dredge-up on mass and metallicity. However, they
conclude that some scaling of their dredge-up law is still
required as their models likely are not able to reproduce directly
either the Galactic or the LMC or SMC C-star luminosity function.
Although much work has been published, no real consensus has been
reached.   The main focus has been on the numerical and physical
treatment of convective boundaries \citep{frost:96,mowlavi:99}.
\cite{herwig:97} and \cite{herwig:99a} showed that even a small
amount of exponential mixing beyond convective boundaries, 
including those of the He-shell flash convection zone, can greatly increase
the model's dredge-up efficiency, maybe even to the extent that
carbon star models of sufficiently low mass can be constructed.
However, it is impossible at this time to know from first
principles how large this convective overshooting really is, and
all implications of this proposition have not yet been evaluated.

Understanding the dredge-up properties of AGB stars is of great
importance for current astrophysical research. Yield predictions
for low mass stars enter models for galactic chemical evolution.
AGB stars serve as diagnostics for extragalactic populations, and
for this purpose the conditions of the O-rich to C-rich
transitions are crucial to know. In fact, C-rich giants are the
brightest infrared population in extra-galactic systems and new
space-based infrared observatories like the Spitzer Space
Telescope emphasize the importance of improved stellar models in
this regard. Finally, the envelope enrichment of AGB stars with
the s-process elements is intimately related to the dredge-up
properties of the models.

We will in the next section describe the results of model calculations
that show the sensitivity of dredge-up predictions to changes in published nuclear
reaction rates, within their uncertainties (\kap{sec:evol}). In \kap{sec:nucrate} we
will present our assesment of recent work on the
$\nvi(\p,\gamma)\ofu$ reaction rate.
Finally we will discuss the results for dredge-up
modelling of AGB stars (\kap{sec:concl}).

\section{Stellar evolution with varying nuclear physics input}
\label{sec:evol}

We  present here the first systematic study based on complete
stellar evolution calculations of how changed nuclear reaction rates
affect the dredge-up and envelope abundance evolution of AGB stars.

Earlier studies have indicated that stronger He-shell flashes are
followed by deeper dredge-up \citep{boothroyd:88}, and that a
decreased energy generation in the H-shell leads to stronger He-shell
flashes \citep{despain:76}. A decrease
in the $\nvi(\p,\gamma)\ofu$ reaction rate leads to a smaller energy
generation, and thereby to a stronger He-shell flash and increased
dredge-up. These earlier findings can be understood qualitatively in
the following way.

A less effective H-shell will cause a slower He-accretion onto the C/O
core, and it will take somewhat longer to reach the He-buffer mass
required to ignite the He-shell and initiate the He-shell flash. In
that case the density will be higher in the He-shell, and thus the
thermonuclear runaway will be larger because the shell is thinner and
partial degeneracy is higher. During H-core burning the published
uncertainties of the $\nvi(\p,\gamma)\ofu$ rate are usually not
influential because the the thermodynamic conditions will adjust
slightly to generate the energy rate required by the stellar mass.  In
the case of shell burning on degenerate cores this is different. The
core is electron-degenerate and it largely determines the
thermodynamic conditions in the H-shell. Thus, a decrease in the
nuclear reaction rate does lead to a smaller He-production
rate, with the consequences of a stronger  He-shell flash and
more efficient dredge-up.

Two reactions are important in the He-shell burning. The
triple-$\alpha$ reaction with its large temperature exponent drives
the thermonuclear runaway of the He-shell. The
$\czw(\alpha,\gamma)\ose$ rate produces oxygen, mainly in the hotter
and deeper layers of the He-shell. The rate uncertainties of both reactions
have been considered.

We have used the stellar evolution code EVOL, which is equipped
with up-to-date input physics \citep[see][for
details]{herwig:03c}. We have calculated seven AGB evolution
tracks from a common starting model with a mass of $2\msun$ and
metallicity $Z=0.01$ at the end of the core He-burning phase. Mass
loss is included according to the formula given by
\cite{bloecker:95a} with a scaling factor $\eta_\mem{BL}=0.1$. All
calculations are evolved until all envelope mass is lost. We
assume exponential, time- and depth-dependent overshooting at the
bottom of the convective envelope ($f_\mem{ov}=0.016$). At the
bottom of the He-shell flash convection zone no overshooting is
allowed in this study.

For the benchmark sequence (ET2) we used the NACRE \citep{angulo:99}
recommended values for all three reactions. In addition, for each
reaction one sequence has been calculated for the lower and for the
upper limit from the NACRE recommendation.  We found that the
uncertainty in the $\czw(\alpha,\gamma)\ose$ reaction has only
marginal influence on the evolution and dredge-up, and we will not
discuss this case any further here. A summary of the remaining five
calculations is given in \tab{tab:sum}.  The factors listed are those
that apply to the analytical reaction rate fits in the temperature
range relevant for H-shell burning ($5\cdot 10^{7}\mem{K}<T<8\cdot
10^{7}\mem{K}$) and He-shell burning ($1\cdot 10^{8}\mem{K}<T<3\cdot
10^{8}\mem{K}$) respectively \citep[see][for details]{herwig:04b}.

The stellar evolution calculations show that the published reaction
rate uncertainties for both the triple-$\alpha$ and the
$\nvi(\p,\gamma)\ofu$ reaction have substantial influence on the
third-dredge-up, and consequently on the formation of carbon stars and
the yields from low-mass AGB stars, as exemplified by carbon.  As
shown in \abb{fig:n14-Mc-lambda} for the $\nvi(\p,\gamma)\ofu$
reaction dredge-up starts at a lower core mass and is larger for a
given core mass if this rate is lower. It is interesting to note that
dredge-up efficiency and derivative quantities (like the yields)
depend in a highly non-linear way on the reaction rate.  For both
reactions the rate uncertainties are individually responsible for the
carbon yield being uncertain by more than a factor of two
(\tab{tab:sum}).  The cases with the higher triple-$\alpha$ rate and
with the lower $\nvi(\p,\gamma)$ rate both lead to carbon star
formation at a lower luminosity (\abb{fig:L-CO}). In both these cases
the stellar model spends five to six thermal pulse (TP) cycles in the
C-rich stage, whereas the other sequences are only able to reach the
C-star regime at about two TPs before the end of the evolution when
the envelope mass is already low.

Modeling of the third dredge-up and comparison with observations is
plagued by a number of theoretical uncertainties, including the
assumptions on mixing processes and mass loss. Modifications like
those proposed with respect to the C/O-ratio dependent molecular
opacities \citep{marigo:02} are important in this regard too.  It is
evident from this differential study that the uncertainties in the two
key nuclear reactions discussed here have a profound impact on the
dredge-up modeling as well.

\section{Rate revision for the $\nvi(\p,\gamma)\ofu$ reaction}
\label{sec:nucrate}
NACRE rates for the $3\alpha$ reaction reflect
present values of the nuclear parameters and seem reliable for the
present temperature range. The situation is different for
$^{14}$N($p,\gamma$)$^{15}$O.

The NACRE tabulation for $^{14}$N($p,\gamma$)$^{15}$O is based
primarily on fits to the data of \citet{schroeder:87}. The
resulting capture reaction to the ground state of $^{15}$O
contributed almost half of the total cross section at low
energies.  This led to a total S factor $S(0)= 3.2 \pm  0.8$
keV\,b and to the NACRE reaction rates \citep{angulo:99}. However,
there was a concern about this result \citep{adelburger:98} because
the fit to the ground state cross section yielded a surprisingly
large value of the gamma width of a (subthreshold) $3/2^+$ state
at $E_x =6.793$ MeV in $^{15}$O, about 7 times the value for the
isospin-analog transition in $^{15}$N.  Such large differences are
rarely, if ever, seen \citep{adelburger:98,brown:04}. A more
detailed reanalysis \citep{angulo:01} of the same data  found that
the ground state transition was small and that $S(0)= 1.77 \pm
0.2$ keV\,b. Direct measurements of the 6.793 state's lifetime
\citep{bertone:01,yamada:04} yielded much smaller gamma widths;
these results and determinations of asymptotic normalization
coefficients (ANC) for the relevant transitions using nuclear
transfer reactions \citep{akram:03} led to a similar conclusion
and $S(0)= 1.7 \pm  0.41$  keV\,b.  All these investigations show
that the ground state transition is small and that the resulting
total S factor is reduced by about a factor of two from the NACRE
result.  Recently, additional $^{14}$N($p,\gamma$)$^{15}$O data
and corrected data from \citet{schroeder:87} were analyzed to
yield $S(0)= 1.70 \pm  0.22$  keV\,b \citep{formicola:03}.  An
independent measurement at TUNL \citep{runkle:04} yielded $S(0)=
1.66 \pm 0.20$ keV\,b.  Because the most accurate results
exist only in preprint form, and because of unresolved issues
involving M1 contributions to the ground state transition
\citep{nelson:03}, we have chosen to use an unweighted average of
the four recent results and a conservative error to obtain $S(0) =
1.70 \pm 0.25$ keV b. This can be expressed as a factor to the
analytical fit of the NACRE rate in the relevant temperature
range, for easy employment in the stellar evolution code, and this
factor is $0.64\pm0.1$.

\section{Discussion and conclusions}
\label{sec:concl} We have shown in \kap{sec:evol} that the
reaction rate uncertainties for the key rates $\nvi(\p,\gamma)$ and
triple-$\alpha$ have significant influence on the dredge-up and yield
predictions of low-mass AGB stars. Within the recommended uncertainties, both a $13\%$ larger
triple-$\alpha$ rate or a $25\%$ lower p-capture rate of $\nvi$ than
adopted by NACRE give \czw\ yield predictions that are higher by a
factor of 2 (\tab{tab:sum}). Preliminary tests indicate
that the superposition of these two uncertainties is non-linear and
leads only to a moderate further increase of the dredge-up and
yields. Clearly a detailed study of the superposition of nuclear
reaction rate uncertainties is desirable. Such studies are
computationally expensive and time consuming, and are therfore
postponed \citep{herwig:04b}. 

 The new $\nvi(\p,\gamma)\ofu$ rate resulting from recent work
 (\kap{sec:nucrate}), is even smaller than the NACRE lower limit used in
 our calculation ET8. Thus, the yields and dredge-up efficiency are
 likely to be somewhat larger in a model with the new recommendation compared
 to those from calculation ET8. In addition, the uncertainty in the
 triple-$\alpha$ reaction has to be factored in as well, allowing a
 range of possible solutions that may extend to still more efficient
 dredge-up. Preliminary analysis of our ongoing calculations suggests
 that a case with the NACRE recommended upper limit of the
 triple-$\alpha$ rate and the lower limit adopted in
 \kap{sec:nucrate} for the $\nvi(\p,\gamma)\ofu$ rate produces about
 3.2 times more carbon than sequence ET2 (both rates NACRE
 recommended).

 Finally, as one of us has argued previously \citep{herwig:99a} 
a small amount of overshooting at bottom of the He-shell flash convection
 zone can accomodate some convincing observational
 constraints related to H-deficient central stars of planetary
 nebulae, which are the progeny of the AGB stars. This overshooting
 might further improve the agreement of models and observations.

We have shown that nuclear reaction rates are an important physics
input when modelling the third dredge-up. We have no indication that
suggests that a dependence of dredge-up on these nuclear reaction
rates is not a universal feature with respect to models of different
mass and metallicity.  

\acknowledgments Research supported in part by the US NSF grants
PHY01-10253 and PHY02-16783, the later  funding the Joint
Institute for Nuclear Astrophysics (JINA), an NSF Physics Frontier
Center.


\clearpage
\begin{figure}
\plotone{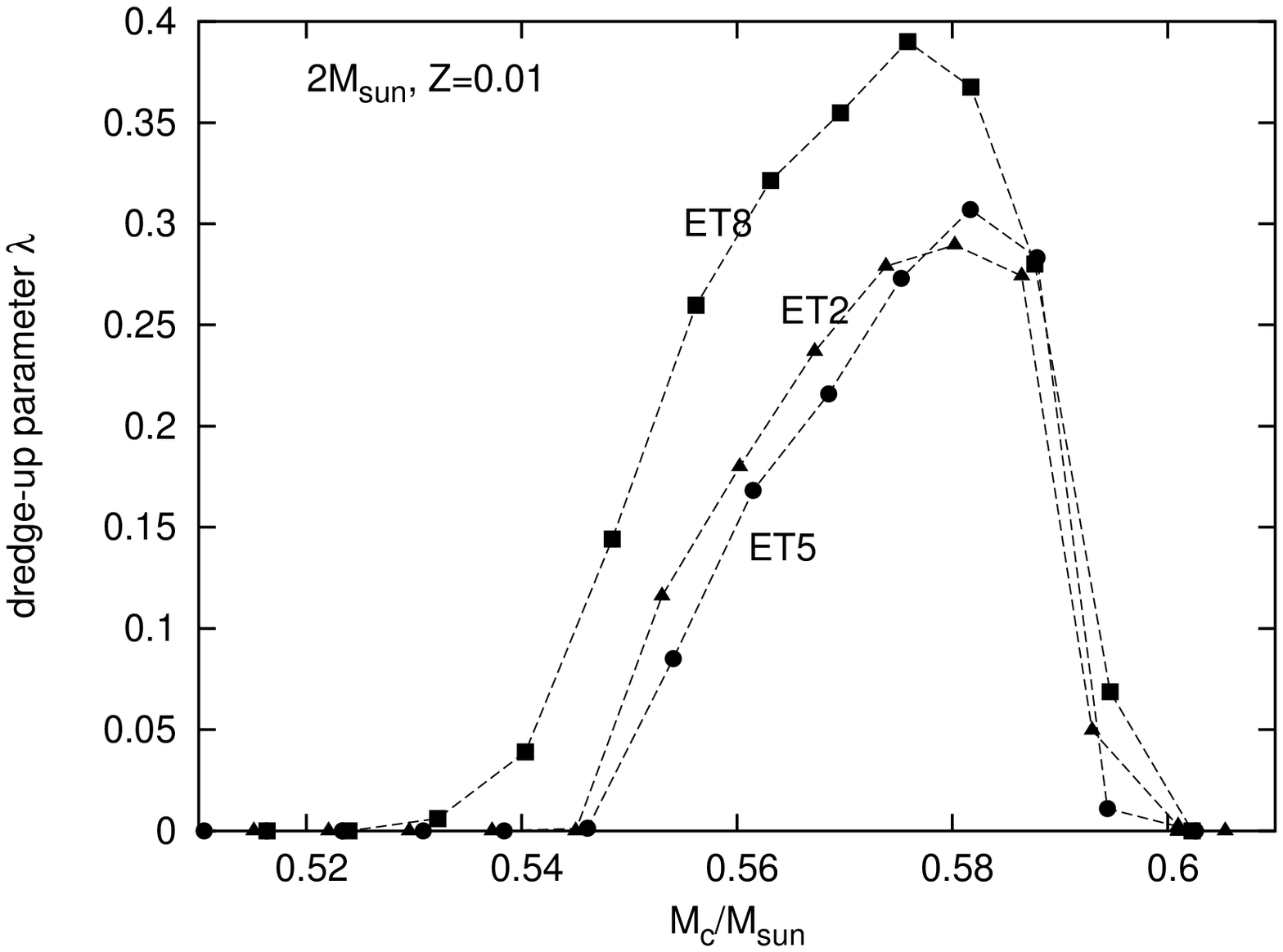}
\caption{Evolution of dredge-up parameter $\lambda$ (ratio of dredged-up mass to
  core mass growth during previous interpulse phase) as a function of
  core mass for the different $\nvi(\p,\gamma)\ofu$ rates.  Labels are
  explained in \tab{tab:sum}. Larger dredge-up is obtained for the
  lower $\nvi + \p$-rate.
  \label{fig:n14-Mc-lambda}}
\end{figure}

\begin{figure}
\plotone{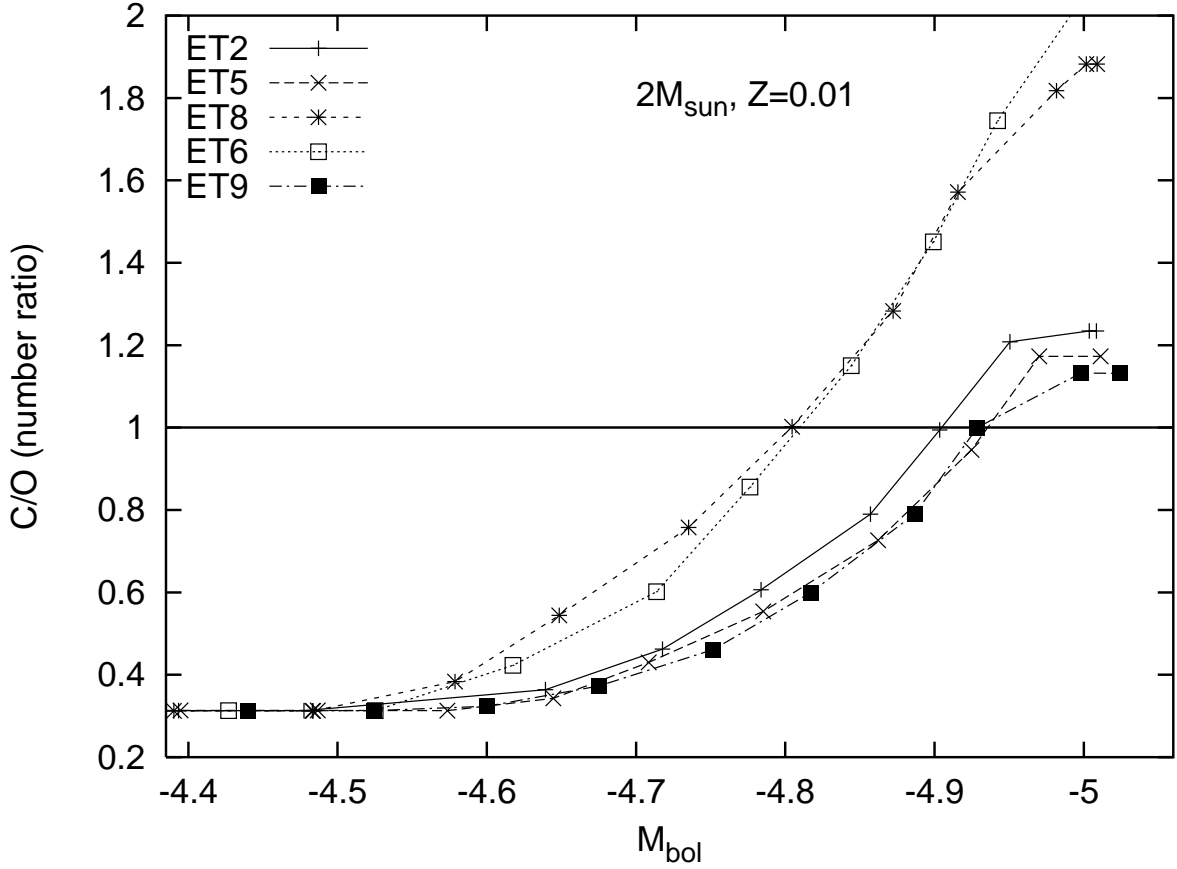}
\caption{Evolution of surface C/O ratio for different assumptions on the nuclear reaction rates
  (see \tab{tab:sum}) as a function of the peak stellar interpulse
  luminosity. The LMC luminosity function peaks at $M_\mem{bol}=-4.9$. The marks
  indicate TPs.
  \label{fig:L-CO}}
\end{figure}

\clearpage
\begin{deluxetable}{cccc}

  \tablecolumns{8} \tablewidth{0pc} \tablecaption{\label{tab:sum}
    Summary of evolution calculations: reaction rate factors and \czw\ yields.}
\tablehead{ \colhead{ID}
    & \colhead{$\nvi(\p,\gamma)\ofu$} & \colhead{triple-$\alpha$} &
    \colhead{$p_{\czw}$$^{a}$} } \startdata
  ET2 & $1.00$& $1.00$ &$ 2.19 \cdot 10^{-3}$\\
  ET5 & $1.33$& $1.00$ &$ 1.95 \cdot 10^{-3}$\\
  ET8 & $0.75$& $1.00$ &$ 4.27 \cdot 10^{-3}$\\
  ET6 & $1.00$& $1.13$ &$ 5.42 \cdot 10^{-3}$\\
  ET9 & $1.00$& $0.82$ &$ 1.79 \cdot 10^{-3}$\\
  \enddata
  \\
  $^{a}$ Carbon yields in \msun: $p_i = -\int_{M_\mem{f}}^{M_\mem{i}}
  (X_i(m)-X_\mem{ini})\, \mem{d}m \punkt$
\end{deluxetable}

 \end{document}